\documentclass{revtex4}
\usepackage{amssymb}
\usepackage{amsfonts}
\usepackage{amsmath}

\input{tcilatex}

\begin{document}

\title{Two-dimensional position-dependent mass Lagrangians;
Superintegrability and exact solvability}
\author{Omar Mustafa}
\email{omar.mustafa@emu.edu.tr}
\affiliation{Department of Physics, Eastern Mediterranean University, G. Magusa, north
Cyprus, Mersin 10 - Turkey,\\
Tel.: +90 392 6301378; fax: +90 3692 365 1604.}

\begin{abstract}
\textbf{Abstract:} The two-dimensional extension of the one-dimensional
PDM-Lagrangians and their nonlocal point transformation mappings into
constant \emph{unit-mass} exactly solvable Lagrangians is introduced. The
conditions on the related two-dimensional Euler-Lagrange equations'
invariance are reported. The mappings from \emph{superintegrable} linear
oscillators into \emph{sub-superintegrable} nonlinear PDM-oscillators are
exemplified by, (i) a \emph{sub-superintegrable} Mathews-Lakshmanan type-I
PDM-oscillator, for a PDM-particle moving in a harmonic oscillator
potential, (ii) a \emph{sub-superintegrable} Mathews-Lakshmanan type-II
PDM-oscillator, for a PDM-particle moving in a constant potential, and (iii)
a \emph{sub-superintegrable} shifted Mathews-Lakshmanan type-III
PDM-oscillator, for a PDM-particle moving in a shifted harmonic oscillator
potential. Moreover, the \emph{superintegrable} shifted linear oscillators
and the isotonic oscillators are mapped into a \emph{sub-superintegrable}
PDM-nonlinear and a \emph{sub-superintegrable} PDM-isotonic oscillators,
respectively.


\end{abstract}

\maketitle

\section{Introduction}

The mathematical challenge associated with the position-dependent mass (PDM)
von Roos Hamiltonian \cite{1}, and the feasible applicability of the PDM
settings in different fields of physics, has inspired a relatively intensive
recent research attention on the quantum mechanical (see the sample of
references \cite{2,3,4,5,6,7,8,9,10,11,12}), classical mechanical and
mathematical (see the sample of references \cite%
{12,13,14,15,16,17,18,19,20,21,22,23,24,25,26,27,28,29,30,31,32,33,34,35,36,37}%
) domains in general. The position-dependent mass is, in principle, a
position-dependent deformation in the standard constant mass settings that
introduces it's own PDM-byproducted reaction-type force $R_{PDM}\left( x,%
\dot{x}\right) =m^{^{\prime }}\left( x\right) \dot{x}^{2}/2$, and manifests
deformation in the potential force field that may inspire nonlocal
space-time point transformations. That is, if a PDM-particle is moving in a
harmonic oscillator potential force field $V\left( x\right) =m\left(
x\right) \omega ^{2}x^{2}/2$, for example, then one may use $q=x\sqrt{%
m\left( x\right) }\Longrightarrow V\left( q\right) =\omega ^{2}q^{2}/2$ to
retain the standard constant mass settings. In the process, some
position-dependent deformation in time may be deemed vital (see \cite{22,38}
for more details on this issue). For some comprehensive discussions on the
quantum mechanical PDM related ordering ambiguity and on the
classical-quantum mechanical correspondence, the reader may refer to (c.f.,,
e.g., \cite{12,17}). However, on the issue of the classical mechanical
equivalence between the Euler-Lagrange's and Newton's equations of motion
one may refer to \cite{22}.

In a very recent study, Mustafa \cite{38} has introduced a general nonlocal
point transformation for one-dimensional PDM Lagrangians and provided their
mappings into a constant \emph{"unit-mass"} Lagrangians in the generalized
coordinates. Therein, it has been shown that the applicability of such
mappings not only results in the linearization of some nonlinear oscillators
but also extends into the extraction of exact solutions of more complicated
dynamical systems. Hereby, the exactly solvable Lagrangians (labeled as 
\emph{"reference/target-Lagrangians"}) are mapped along with their exact
solutions into PDM-Lagrangians (labeled as \emph{%
"target/reference-Lagrangians"}). It would be natural and interesting to
extend/generalize Mustafa's methodical proposal \cite{38} to deal with
Lagrangians in more than one-dimension. Therefore, the current methodical
proposal is a parallel extension to \cite{38} and deals with PDM Lagrangians
in two-dimensions.

However, in handling such higher dimensional Lagrangians the notion/concept
of \emph{superintegrability} (c.f., e.g., \cite%
{39,40,41,42,43,44,45,46,47,48,49,50,51} and related references cited
therein) is unavoidable. A Lagrangian system is said to be \emph{%
superintegrable} if it admits the Liouville-Arnold sense of integrability
and introduces more constants of motion (also called integrals of motion)
than the degrees of freedom the system is moving within (c.f., e.g., \cite%
{45}). The set of the two-dimensional isotonic oscillator potentials%
\begin{equation}
V(x_{1},x_{2})=\frac{1}{2}\left( \omega _{1}^{2}x_{1}^{2}+\omega
_{2}^{2}x_{2}^{2}+\frac{\beta _{1}}{x_{1}^{2}}+\frac{\beta _{2}}{x_{2}^{2}}%
\right) ,
\end{equation}%
for example (a more general extension of the Smorodinsky-Winternitz isotonic
oscillator potentials where $\ \omega _{1}=\omega _{2}=\omega $) is known to
be \emph{superintegrable} (c.f., e.g., \cite{39,40,41,42,46,47}).\ The
details on their \emph{superintegrability} (c.f.,e.g., \cite{39,41}) or
their \emph{maximal superintegrability} (c.f., e.g., \cite{40})
classification criteria lay far beyond the scope of our study here. They are
considered as \emph{superintegrable} potentials throughout the current
methodical proposal, though we verify their \emph{superintegrability} in
brief to make the current methodical proposal self-contained. On the other
hand, under some nonlocal space-time transformation (see (10) below) the
two-dimensional \emph{superintegrable} isotonic oscillator potentials%
\begin{equation}
V(q_{1},q_{2})=\frac{1}{2}\left( \omega _{1}^{2}q_{1}^{2}+\omega
_{2}^{2}q_{2}^{2}+\frac{\beta _{1}}{q_{1}^{2}}+\frac{\beta _{2}}{q_{2}^{2}}%
\right) ,
\end{equation}%
(where $q_{j}\equiv q_{j}\left( x_{j}\right) \,;j=1,2$ are some invertible, $%
\partial x_{j}/\partial q_{j}\neq 0\neq \partial q_{j}/\partial x_{j}$,
generalized coordinates) may yield a PDM-deformed isotonic oscillator
potentials. Hereby,\ we argue that, if the exactly solvable \emph{%
"superintegrable reference-Lagrangians"} are mapped into\ \emph{"PDM
target-Lagrangians"}, then the set of \emph{"PDM target-Lagrangians"} is a
set of \emph{"sub-superintegrable PDM-Lagrangians"} (as shall be so labeled
hereinafter). In addition to our main objective to extend/generalize
Mustafa's methodical proposal \cite{38} to deal with two-dimensional
Lagrangians, we anticipate that the introduction of the terminology of \emph{%
"sub-superintegrability"} \ would\ yet add a new flavour/concept to \emph{%
"superintegrability"}. The organization of our methodical proposal is in
order.

The two-dimensional nonlocal space-time PDM transformation and the
Euler-Lagrange equations' invariance are discussed in section 2. Therein,
one would observe that the obvious non-separability of the two-dimensional 
\emph{"PDM target-Lagrangians"} $L\left( x_{1},x_{2},\dot{x}_{1},\dot{x}%
_{2};t\right) $ is accompanied by the separability of the two-dimensional 
\emph{"reference-Lagrangians"} $L\left( q_{1},q_{2},\tilde{q}_{1},\tilde{q}%
_{2};\tau \right) $. Mandating in effect \emph{"sub-separability"} of the
two-dimensional \emph{"PDM target-Lagrangians"} $L\left( x_{1},x_{2},\dot{x}%
_{1},\dot{x}_{2};t\right) $, as a result of the two-dimensional nonlocal PDM
transformations. In section 3, the mapping from the \emph{superintegrability}
of some linear oscillators into \emph{sub-superintegrability} of nonlinear
PDM-oscillators is introduced. To make the current methodical proposal
self-contained, we discuss in short the superintegrability of the \emph{%
reference-Lagrangians} (although similar discussions are available in the
literature). Three two-dimensional \emph{sub-superintegrable}
Mathews-Lakshmanan type PDM-oscillators are used for clarification. They
are, (i) a \emph{sub-superintegrable} Mathews-Lakshmanan type-I
PDM-oscillator for a PDM-particle moving in a harmonic oscillator potential,
(ii) a \emph{sub-superintegrable} Mathews-Lakshmanan type-II PDM-oscillator
for a PDM-particle moving in a constant potential, (iii) a \emph{%
sub-superintegrable} shifted Mathews-Lakshmanan type-III PDM-oscillator for
a PDM-particle moving in a a shifted harmonic oscillator potential. In the
same section, we discuss (in short) the superintegrability of some shifted
linear oscillators (a new superintegrable model to the best of our
knowledge) and report their mapping into sub-superintegrable nonlinear
PDM-Oscillators. We use, moreover, a \emph{superintegrable} isotonic
oscillator and map it into a \emph{sub-superintegrable} PDM isotonic
oscillator. Of course the mappings also include exact solvability of the 
\emph{"reference-Lagrangians"} at hand as well. Our concluding remarks are
given in section 4.

\section{Two-dimensional nonlocal PDM-point transformations and
Euler-Lagrange's invariance}

The Lagrangian of a particle with a constant \emph{"unit mass"} moving in
the generalized coordinates $\left( q_{1},q_{2}\right) \equiv \left(
q_{1}\left( x_{1}\right) ,q_{2}\left( x_{2}\right) \right) $, under the
influence of a potential force field $V(q_{1},q_{2}),$ and a
deformed/re-scaled time $\tau $, is given by%
\begin{equation}
L\left( q_{1},q_{2},\tilde{q}_{1},\tilde{q}_{2};\tau \right) =\frac{1}{2}%
\left( \tilde{q}_{1}^{2}+\tilde{q}_{2}^{2}\right) -V(q_{1},q_{2})\text{ };%
\text{ \ }\tilde{q}_{j}=\frac{dq_{j}}{d\tau };\,j=1,2.
\end{equation}%
The\ corresponding Euler-Lagrange equations%
\begin{equation}
\frac{d}{d\tau }\left( \frac{\partial L}{\partial \tilde{q}_{j}}\right) -%
\frac{\partial L}{\partial q_{j}}=0;\,\,j=1,2,
\end{equation}%
therefore, read 
\begin{equation}
\text{\ }\frac{d}{d\tau }\tilde{q}_{j}+\frac{\partial }{\partial q_{j}}%
V(q_{1},q_{2})=0;\,j=1,2.
\end{equation}%
One should notice that for a \emph{"unit mass"} particle moving in a \emph{%
free} force field $V(q_{1},q_{2})=0\Longrightarrow d\tilde{q}_{j}\left(
x_{j}\right) /d\tau =0$, hence the linear momenta $\tilde{q}_{1}\left(
x_{1}\right) $ and $\tilde{q}_{2}\left( x_{2}\right) $ are conserved
quantities ( in this particular case) and serve as fundamental integrals
(i.e., constants of motion). However, for the set of potentials in the from
of 
\begin{equation}
V(q_{1},q_{2})=V_{1}\left( q_{1}\right) +V_{2}\left( q_{2}\right) \neq
0;\,V_{j}\left( q_{j}\right) \neq 0;\,j=1,2
\end{equation}%
(which is the set of potential force fields of our interest in the current
study) one may recast (5) as%
\begin{equation}
\text{\ }\frac{d}{d\tau }\tilde{q}_{j}\left( x_{j}\right) +\frac{\partial }{%
\partial q_{j}}V_{j}(q_{j})=0\Longrightarrow \tilde{q}_{j}\left( x\right) 
\text{\ }\frac{d}{d\tau }\tilde{q}_{j}\left( x\right) +\tilde{q}_{j}\left(
x\right) \frac{\partial }{\partial q_{j}}V_{j}(q_{j})=0.
\end{equation}%
to obtain two integrals of motion $I_{1}=E_{1}$ and $I_{2}=E_{2}$ via 
\begin{equation}
\frac{d}{d\tau }\left[ \frac{1}{2}\tilde{q}_{j}^{2}+V_{j}(q_{j})\right]
=0\Longrightarrow \frac{dE_{j}}{d\tau }=\frac{dI_{j}}{d\tau }=0;\,j=1,2.
\end{equation}%
Yet, the conservation of the total energy $E_{tot}$ is a natural and an
immediate consequence of such settings. That is,%
\begin{equation}
\frac{d}{d\tau }\left[ E_{1}+E_{2}\right] =\frac{d}{d\tau }\left[ \frac{1}{2}%
\left( \tilde{q}_{1}^{2}+\tilde{q}_{2}^{2}\right) +V_{1}\left( q_{1}\right)
+V_{2}\left( q_{2}\right) \right] =0\Longrightarrow \frac{dE_{tot}}{d\tau }=0
\end{equation}%
The separability and/or integrability of the above system is obvious,
therefore.

Next, let us introduce the nonlocal point transformation of the form%
\begin{equation}
q_{j}\equiv q_{j}\left( x_{j}\right) =\int \sqrt{g\left( \bar{x}\right) }%
dx_{j}\text{, \ }\tau =\int f\left( \bar{x}\right) dt\Longrightarrow \frac{%
d\tau }{dt}=f\left( \bar{x}\right) \neq 0\text{ };\text{ \ }x_{j}\equiv
x_{j}\left( t\right) ,\,\bar{x}=x_{1},x_{2}.
\end{equation}%
Consequently, with an overhead dot to identify total time $t$ derivative, 
\begin{equation}
\frac{dq_{j}}{d\tau }=\tilde{q}_{j}=\frac{\dot{x}_{j}\sqrt{g\left( \bar{x}%
\right) }}{f\left( \bar{x}\right) }\text{ },\text{ \ \ \ }\frac{d}{d\tau }%
\tilde{q}_{j}\left( x_{j}\right) =\frac{\sqrt{g\left( \bar{x}\right) }}{%
f\left( \bar{x}\right) ^{2}}\left[ \ddot{x}_{j}+\left( \frac{1}{2}\frac{\dot{%
g}\left( \bar{x}\right) }{g\left( \bar{x}\right) }-\frac{\dot{f}\left( \bar{x%
}\right) }{f\left( \bar{x}\right) }\right) \dot{x}_{j}\right] .
\end{equation}%
Which when substituted in (7) would , in a straightforward manner, result in 
\begin{equation}
\left[ \dot{x}_{1}\,\ddot{x}_{1}+\dot{x}_{2}\,\ddot{x}_{2}+\left( \frac{1}{2}%
\frac{\dot{g}\left( \bar{x}\right) }{g\left( \bar{x}\right) }-\frac{\dot{f}%
\left( \bar{x}\right) }{f\left( \bar{x}\right) }\right) \left( \dot{x}%
_{1}^{2}+\dot{x}_{2}^{2}\right) \right] +\frac{f\left( \bar{x}\right) ^{2}}{%
g\left( \bar{x}\right) }\frac{d}{dt}V\left( \bar{x}\right) =0;\,x_{j}\equiv
x_{j}\left( q_{j}\right) ,
\end{equation}%
where%
\begin{equation*}
\frac{d}{dt}V\left( \bar{x}\right) =\left( \dot{x}_{1}\frac{\partial }{%
\partial x_{1}}V_{1}\left( x_{1}\right) +\dot{x}_{2}\frac{\partial }{%
\partial x_{2}}V_{2}\left( x_{2}\right) \right) .
\end{equation*}

On the other hand, for a two-dimensional position-dependent mass particle
moving in a force field $V\left( x_{1},x_{2}\right) =V_{1}\left(
x_{1}\right) +V_{2}\left( x_{2}\right) $ the Lagrangian%
\begin{equation}
L\left( x_{1},x_{2},\dot{x}_{1},\dot{x}_{2};t\right) =\frac{1}{2}m\left( 
\bar{x}\right) \left( \dot{x}_{1}^{2}+\dot{x}_{2}^{2}\right) -\left[
V_{1}\left( x_{1}\right) +V_{2}\left( x_{2}\right) \right] ;,
\end{equation}%
in the Cartesian coordinates, would yield two Euler-Lagrange equations 
\begin{equation}
m\left( \bar{x}\right) \ddot{x}_{j}+\dot{m}\left( \bar{x}\right) \dot{x}_{j}-%
\frac{1}{2}\frac{\partial m\left( \bar{x}\right) }{\partial x_{j}}\left( 
\dot{x}_{1}^{2}+\dot{x}_{2}^{2}\right) +\frac{\partial }{\partial x_{j}}%
V_{j}\left( x_{j}\right) =0;\,j=1,2.
\end{equation}%
The non-separability of this system is obviously manifested by the
position-dependent mass term. However, when multiplied, from the left, by $%
\dot{x}_{j}$ it reads 
\begin{equation}
m\left( \bar{x}\right) \dot{x}_{j}\ddot{x}_{j}+\dot{m}\left( \bar{x}\right) 
\dot{x}_{j}^{2}-\frac{1}{2}\left( \dot{x}_{j}\frac{\partial m\left( \bar{x}%
\right) }{\partial x_{j}}\right) \left[ \dot{x}_{1}^{2}+\dot{x}_{2}^{2}%
\right] +\left( \dot{x}_{j}\frac{\partial }{\partial x_{j}}\right)
V_{j}\left( x_{j}\right) =0;\,j=1,2.
\end{equation}%
Consequently, the addition of the two equations of (15) yields%
\begin{equation}
\dot{x}_{1}\ddot{x}_{1}+\dot{x}_{2}\ddot{x}_{2}+\frac{1}{2}\frac{\dot{m}%
\left( \bar{x}\right) }{m\left( \bar{x}\right) }\left( \dot{x}_{1}^{2}+\dot{x%
}_{2}^{2}\right) +\frac{1}{m\left( \bar{x}\right) }\frac{d}{dt}V\left( \bar{x%
}\right) =0,
\end{equation}%
and hence%
\begin{equation}
\frac{d}{dt}\left[ \frac{1}{2}m\left( \bar{x}\right) \,\left( \dot{x}%
_{1}^{2}+\dot{x}_{2}^{2}\right) +V_{1}\left( x_{1}\right) +V_{2}\left(
x_{2}\right) \right] =0\Longrightarrow \frac{dE_{tot}}{dt}=0.
\end{equation}%
Hereabout, we emphasize that the conservation of the total energy $E_{tot}$
in (17) can \emph{never} be considered as an immediate consequence of the
sum of two fundamental integrals of motion $E_{x_{1}}$ and $E_{x_{2}}$. The
time evolutions of $E_{x_{j}j}$'s do not satisfy (15). i.e., one may easily
show that 
\begin{equation}
\frac{d}{dt}E_{x_{j}}=\frac{d}{dt}\left[ \frac{1}{2}m\left( \bar{x}\right) 
\dot{x}_{j}^{2}+V_{j}\left( x_{j}\right) \right] =m\left( \bar{x}\right) 
\dot{x}_{j}\ddot{x}_{j}+\frac{1}{2}\dot{m}\left( \bar{x}\right) \dot{x}%
_{j}^{2}+\dot{x}_{j}\frac{\partial }{\partial x_{j}}V_{j}\left( x_{j}\right)
\neq 0
\end{equation}%
compared to (15). The third term in (15) is missing in (18). Moreover, the
comparison between (12) and (16) obviously suggests that the Euler-Lagrange
equations of motion (12) and (16) are identical if and only if $f\left( \bar{%
x}\right) $ and $g\left( \bar{x}\right) $ satisfy the conditions 
\begin{equation}
g\left( \bar{x}\right) =m\left( \bar{x}\right) \,f\left( \bar{x}\right)
^{2}\Longleftrightarrow \frac{1}{2}\frac{\dot{g}\left( \bar{x}\right) }{%
g\left( \bar{x}\right) }-\frac{\dot{f}\left( \bar{x}\right) }{f\left( \bar{x}%
\right) }=\frac{1}{2}\frac{\dot{m}\left( \bar{x}\right) }{m\left( \bar{x}%
\right) }\Longleftrightarrow q_{j}=\int \sqrt{m\left( \bar{x}\right) }%
f\left( \bar{x}\right) dx_{j}.
\end{equation}%
As such, it is clear that the functional nature/structure of the
position-dependent mass $m\left( \bar{x}\right) $ determines the
nature/structure of the nonlocal transformation functions $g\left( \bar{x}%
\right) $ and $\,f\left( \bar{x}\right) .$ For the sake of simplicity,
however, we shall work with $m\left( \bar{x}\right) \equiv m\left( r\right)
\Longrightarrow g\left( \bar{x}\right) \equiv g\left( r\right) $ and $%
\,f\left( \bar{x}\right) \equiv $ $f\left( r\right) $. Hence, $m\left(
r\right) $, $g\left( r\right) $ and $f\left( r\right) $ are well behaved
functions of explicit dependence on $r=\sqrt{x_{1}^{2}+x_{2}^{2}}$, if not
otherwise mentioned.

At this point, one may safely conclude that under such invertible (i.e., the
Jacobian determinant $det\left( \partial x_{i}/\partial q_{i}\right) \neq 0$%
) nonlocal transformation, (10) and (19), the two-dimensional Euler-Lagrange
equations, (12) and (16), remain invariant. That is, the Lagrangian $L\left(
q_{1},q_{2},\tilde{q}_{1},\tilde{q}_{2};\tau \right) ,$ in (3), non-locally
transforms (via (10) and (19)) into the PDM Lagrangian $L\left( x_{1},x_{2},%
\dot{x}_{1},\dot{x}_{2};t\right) $, in (13), and leaves in the process the
corresponding two-dimensional Euler-Lagrange equations of motions, (12) and
(16), invariant. The mapping%
\begin{equation}
L\left( q_{1},q_{2},\tilde{q}_{1},\tilde{q}_{2};\tau \right)
\Longleftrightarrow \left\{ 
\begin{array}{c}
g\left( r\right) =m\left( r\right) f\left( r\right) ^{2}\medskip \medskip \\ 
q_{j}=\int \sqrt{m\left( r\right) }f\left( r\right)
dx_{j}\,;\,j=1,2\smallskip \medskip \\ 
\tau =\int f\left( r\right) dt\medskip \\ 
\tilde{q}_{j}=\dot{x}_{j}\sqrt{m\left( r\right) };\,j=1,2%
\end{array}%
\right\} \Longleftrightarrow L\left( x_{1},x_{2},\dot{x}_{1},\dot{x}%
_{2};t\right) ,
\end{equation}%
between the \emph{"unit mass" } Lagrangian $L\left( q_{1},q_{2},\tilde{q}%
_{1},\tilde{q}_{2};\tau \right) $ and PDM Lagrangian $L\left( x_{1},x_{2},%
\dot{x}_{1},\dot{x}_{2};t\right) $ is clear, therefore.

Nevertheless, the obvious non-separability of the two-dimensional
Euler-Lagrange equations associated with the PDM Lagrangian $L\left(
x_{1},x_{2},\dot{x}_{1},\dot{x}_{2};t\right) $ (hence, $L\left( x_{1},x_{2},%
\dot{x}_{1},\dot{x}_{2};t\right) $ is non-separable) is accompanied by the
separability of the two-dimensional Euler-Lagrange equations associated with
unit mass Lagrangian $L\left( q_{1},q_{2},\tilde{q}_{1},\tilde{q}_{2};\tau
\right) $ (hence, $L\left( q_{1},q_{2},\tilde{q}_{1},\tilde{q}_{2};\tau
\right) $ is separable). This should, in turn, mandate the notion of
"sub-separability" of the two-dimensional Lagrangian $L\left( x_{1},x_{2},%
\dot{x}_{1},\dot{x}_{2};t\right) $ as a result of the two-dimensional PDM
nonlocal transformations in (20). Likewise, if the two-dimensional unit mass
Lagrangian $L\left( q_{1},q_{2},\tilde{q}_{1},\tilde{q}_{2};\tau \right) $
admits superseparability and/or superintegrability then the two-dimensional
PDM Lagrangian $L\left( x_{1},x_{2},\dot{x}_{1},\dot{x}_{2};t\right) $ may
very well be labeled as sub-superseparable and/or sub-superintegrable. The
latter forms the focal point of our study here and shall be clarified in the
forthcoming experimental examples.

\section{From superintegrability to sub-superintegrability; two-dimensional
PDM-oscillators}

Although the superintegrability of some of the \emph{reference
superintegrable} oscillators, we use here, is verified in the literature, we
recycle them in such a way that serves our methodical proposal and keeps it
self-contained. The\emph{\ superintegrable linear oscillators}, the \emph{%
superintegrable shifted-oscillators} (is a new \emph{superintegrable}
oscillator model, to the best of our knowledge), and the \emph{%
superintegrable isotonic oscillators} are used here as illustrative
examples. They are mapped along with their exact solutions into \emph{%
sub-superintegrable PDM-oscillators}.

\subsection{Superintegrable linear oscillators into sub-superintegrable
nonlinear PDM-oscillators}

Consider a \emph{unit mass }particle moving under the influence of the
two-dimensional oscillators potential 
\begin{equation}
V(q_{1},q_{2})=\frac{1}{2}\left( \omega _{1}^{2}q_{1}^{2}+\omega
_{2}^{2}q_{2}^{2}\right) \Longrightarrow V_{j}\left( q_{j}\right) =\frac{1}{2%
}\omega _{j}^{2}q_{j}^{2},\,\omega _{j}=n_{j}\omega _{\circ };\,j=1,2,
\end{equation}%
in the generalized coordinates. Then, the corresponding two-dimensional
Lagrangian%
\begin{equation}
L\left( q_{1},q_{2},\tilde{q}_{1},\tilde{q}_{2};\tau \right) =\frac{1}{2}%
\left( \tilde{q}_{1}^{2}+\tilde{q}_{2}^{2}\right) -\frac{1}{2}\left( \omega
_{1}^{2}q_{1}^{2}+\omega _{2}^{2}q_{2}^{2}\right) ,
\end{equation}%
leads to Euler-Lagrange equations%
\begin{equation}
\frac{d}{d\tau }\tilde{q}_{j}+\omega _{j}^{2}q_{j}=0\Longrightarrow
q_{j}\left( \tau \right) =A_{j}\cos \left( \omega _{j}\tau +\varphi \right)
,\,\,j=1,2,
\end{equation}%
subjected to the boundary conditions $\,q_{j}\left( 0\right) =A_{j},\,\tilde{%
q}_{j}\left( 0\right) =0$, say. It obviously admits separability and is
known to satisfy the superintegrability conditions (c.f. e.g., \cite%
{39,40,41,42}) via the use of \ a complex factorization technique (c.f.,
e.g., \cite{43,44}). That is, if we introduce the two complex functions 
\begin{equation}
Q_{j}=\tilde{q}_{j}+i\omega _{j}q_{j}\Longrightarrow \frac{d}{d\tau }%
Q_{j}=i\omega _{j}Q_{j};\,\,j=1,2,
\end{equation}%
then the functions%
\begin{equation}
Q_{jk}=Q_{j}^{\omega _{k}}\left( Q_{k}^{\ast }\right) ^{\omega
_{j}};\,\,j,k=1,2,
\end{equation}%
represent \emph{complex }constants of motion with vanishing
deformed/rescaled-time evolution $\tau $,%
\begin{equation}
\frac{d}{d\tau }Q_{jk}=i\left( \omega _{j}\omega _{k}-\omega _{k}\omega
_{j}\right) Q_{jk}=0.
\end{equation}%
Moreover, one can, in a straightforward manner, verify that 
\begin{equation}
Q_{jj}=\tilde{q}_{j}^{2}+\omega _{j}^{2}q_{j}^{2}\Longrightarrow
2E_{1}=I_{1}=Q_{11},\,2E_{2}=I_{2}=Q_{22},
\end{equation}%
where $I_{1}$ and $I_{2}$ are two fundamental integrals of motion. Yet, for
the isotropic oscillator $\omega _{1}=\omega _{2}=\omega _{\circ }$, for
example,%
\begin{equation}
Q_{12}=\left( \tilde{q}_{1}\tilde{q}_{2}+\omega _{\circ
}^{2}q_{1}q_{2}\right) +i\omega _{\circ }\left( q_{1}\tilde{q}_{2}-q_{2}%
\tilde{q}_{1}\right) =I_{3}+iI_{4}
\end{equation}%
identifies two more integrals of motion 
\begin{equation}
I_{3}=\func{Re}Q_{12}=\tilde{q}_{1}\tilde{q}_{2}+\omega _{\circ
}^{2}q_{1}q_{2}\text{ \ ; \ \ }I_{4}=\func{Im}Q_{12}=\omega _{\circ }\left(
q_{1}\tilde{q}_{2}-q_{2}\tilde{q}_{1}\right) .
\end{equation}%
which are, in the general case $\omega _{1}\neq \omega _{2}$, polynomials in
the momenta. Therefore, our two-dimensional Lagrangian (22) admits
superintegrability in the generalized coordinates $\left( q_{1},q_{2}\right) 
$ and in the deformed/rescaled time $\tau $. The details on such
superintegrability are far beyond the scope of our the current methodical
proposal, though can be traced through the comprehensive article of Ranada 
\cite{42} and related references cited therein.

\subsubsection{Sub-superintegrable Mathews-Lakshmanan type-I PDM-oscillators}

Let us now consider a PDM particle $m\left( r\right) $\ moving in the
harmonic oscillator force field 
\begin{equation*}
V\left( x_{1},x_{2}\right) =\frac{1}{2}m\left( r\right) \omega ^{2}\left(
x_{1}^{2}+x_{2}^{2}\right) ,
\end{equation*}%
with the corresponding PDM Lagrangian%
\begin{equation}
L\left( x_{1},x_{2},\dot{x}_{1},\dot{x}_{2};t\right) =\frac{1}{2}m\left(
r\right) \left[ \dot{x}_{1}^{2}+\dot{x}_{2}^{2}-\omega ^{2}\left(
x_{1}^{2}+x_{2}^{2}\right) \right] .
\end{equation}%
This Lagrangian indulges one and only one integral offered by the total
energy $E_{tot}$ as in (17), and the corresponding Euler-Lagrange equations
are non-separable. However, with the substitution%
\begin{equation}
q_{j}=x_{j}\sqrt{m\left( r\right) }\,;\smallskip \medskip \,j=1,2,
\end{equation}%
one can non-locally transform $L\left( x_{1},x_{2},\dot{x}_{1},\dot{x}%
_{2};t\right) $ of (30) into $L\left( q_{1},q_{2},\tilde{q}_{1},\tilde{q}%
_{2};\tau \right) $ of (22). Moreover, if this substitution is used along
with the transformation in (20) we get%
\begin{equation}
q_{j}=x_{j}\sqrt{m\left( r\right) }\Longrightarrow \frac{dq_{j}}{dx_{j}}=%
\sqrt{m\left( r\right) }\left[ 1+\frac{m^{\prime }\left( r\right) }{2m\left(
r\right) }\left( \frac{x_{j}^{2}}{r}\right) \right] ,\medskip
\end{equation}%
and%
\begin{equation*}
q_{j}=\int \sqrt{m\left( r\right) }f\left( r\right) dx_{j}\Longrightarrow 
\frac{dq_{1}}{dx_{1}}+\frac{dq_{2}}{dx_{2}}=2\sqrt{m\left( r\right) }f\left(
r\right) .
\end{equation*}%
hence%
\begin{equation}
f\left( r\right) =1+\frac{1}{4}\frac{m^{^{\prime }}\left( r\right) }{m\left(
r\right) }r.
\end{equation}%
At this point, one should be aware that we are interested in $m\left(
r\right) $ and $f\left( r\right) $ that are only explicit function in $r=%
\sqrt{x_{1}^{2}+x_{2}^{2}}$. Obviously, moreover, the choice of $f\left(
r\right) $ would determine the position-dependent mass function $m\left(
r\right) $ (of course the other way around works as well). This is clarified
in the following assumption. Let us assume that%
\begin{equation}
f\left( r\right) =m\left( r\right) -\frac{1}{4}\frac{m^{^{\prime }}\left(
r\right) }{m\left( r\right) }r
\end{equation}%
to obtain a position-dependent mass of the form%
\begin{equation}
f\left( r\right) =m\left( r\right) -\frac{1}{4}\frac{m^{^{\prime }}\left(
r\right) }{m\left( r\right) }r=1+\frac{1}{4}\frac{m^{^{\prime }}\left(
r\right) }{m\left( r\right) }r\Longleftrightarrow m\left( r\right) =\frac{1}{%
1\pm \beta r^{2}};\,\beta \geq 0.
\end{equation}%
Then the corresponding two-dimensional PDM Lagrangian of (30) reads a
two-dimensional Mathews-Lakshmanan type-I oscillator%
\begin{equation}
L\left( x_{1},x_{2},\dot{x}_{1},\dot{x}_{2};t\right) =\frac{1}{2}\frac{\left[
\dot{x}_{1}^{2}+\dot{x}_{2}^{2}-\omega ^{2}\left( x_{1}^{2}+x_{2}^{2}\right) %
\right] }{1\pm \beta \left( x_{1}^{2}+x_{2}^{2}\right) }.
\end{equation}

A Lagrangian of this type is neither separable nor superintegrable.
Nevertheless, our two-dimensional PDM Lagrangian $L\left( x_{1},x_{2},\dot{x}%
_{1},\dot{x}_{2};t\right) $ (36) nonlocaly transforms into a \emph{%
superintegrable} two-dimensional Lagrangian $L\left( q_{1},q_{2},\tilde{q}%
_{1},\tilde{q}_{2};\tau \right) $ (22). Hence the two-dimensional PDM\
Lagrangian $L\left( x_{1},x_{2},\dot{x}_{1},\dot{x}_{2};t\right) $ of (36)
is a \emph{sub-superintegrable \ }PDM Lagrangian and the corresponding
Euler-Lagrange equations (16) admit exact solutions%
\begin{equation}
x_{j}\left( t\right) =A_{j}\cos \left( \Omega t+\varphi \right) ;\,\Omega
^{2}=\frac{\omega ^{2}}{1\pm \beta \left( A_{1}^{2}+A_{2}^{2}\right) }.
\end{equation}%
Likewise, the reversed process is equally valid. That is, the relation 
\begin{equation}
\left\{ 
\begin{tabular}{c}
\emph{\ Superintegrable}\medskip \\ 
$L\left( q_{1},q_{2},\tilde{q}_{1},\tilde{q}_{2};\tau \right) \medskip $ \\ 
of \ (22) with\medskip \\ 
$q_{j}\left( \tau \right) =A_{j}\cos \left( \omega \tau +\varphi \right) $%
\end{tabular}%
\right\} \Longleftrightarrow \left\{ 
\begin{tabular}{l}
$\text{Nonlocal transformation\medskip }$ \\ 
$q_{j}=x_{j}\sqrt{m\left( r\right) }\,;\smallskip \medskip \,j=1,2,$ \\ 
$\tilde{q}_{j}=\dot{x}_{j}\sqrt{m\left( r\right) }\medskip $ \\ 
$f\left( r\right) =m\left( r\right) -\frac{1}{4}\frac{m^{^{\prime }}\left(
r\right) }{m\left( r\right) }r\medskip $ \\ 
$m\left( r\right) =1/\left( 1\pm \beta r^{2}\right) $%
\end{tabular}%
\right\} \Longleftrightarrow \left\{ 
\begin{tabular}{c}
\emph{Sub-superintegrable}\medskip \\ 
$L\left( x_{1},x_{2},\dot{x}_{1},\dot{x}_{2};t\right) \medskip $ \\ 
of \ (36) with\medskip \\ 
$x_{j}\left( t\right) =A_{j}\cos \left( \Omega t+\varphi \right) \,\medskip $
\\ 
$\text{ \ }\Omega ^{2}=\frac{\omega ^{2}}{1\pm \beta \left(
A_{1}^{2}+A_{2}^{2}\right) }\medskip $%
\end{tabular}%
\right\} .
\end{equation}%
provides the exact mapping from the \emph{superintegrability} and exact
solvability of (22) into the \emph{sub-superintegrability} and exact
solvability of the PDM Mathews-Lakshmanan type-I Lagrangian (36).

\subsubsection{Sub-superintegrable Mathews-Lakshmanan type-II PDM-oscillators%
}

Consider a PDM particle $m\left( r\right) $ moving in a constant potential
force field of the form%
\begin{equation}
V\left( x_{1},x_{2}\right) =\frac{1}{2}m\left( r\right) \omega ^{2}\left(
\xi _{1}^{2}+\xi _{2}^{2}\right)
\end{equation}%
with the corresponding Lagrangian%
\begin{equation}
L\left( x_{1},x_{2},\dot{x}_{1},\dot{x}_{2};t\right) =\frac{1}{2}m\left(
r\right) \left[ \dot{x}_{1}^{2}+\dot{x}_{2}^{2}-\omega ^{2}\left( \xi
_{1}^{2}+\xi _{2}^{2}\right) \right] .
\end{equation}%
where $\xi _{1},$ $\xi _{2}\in 
\mathbb{R}
$ are constants. This Lagrangian has the total energy $E_{tot}$ of (17) as
the only integral of motion and the corresponding Euler-Lagrange equations
are non-separable. However, the substitution of $q_{j}=\xi _{j}\sqrt{m\left(
r\right) }$ would nonlocaly transform it into the \emph{superintegrable}
Lagrangian $L\left( q_{1},q_{2},\tilde{q}_{1},\tilde{q}_{2};\tau \right) $
of (22) for $\omega _{1}=\omega _{2}=\omega $. Under such settings,%
\begin{equation}
q_{j}=\int \sqrt{m\left( r\right) }f\left( r\right) dx_{j}=\xi _{j}\sqrt{%
m\left( r\right) }\Longrightarrow f\left( r\right) =\frac{\xi _{j}}{2}\frac{%
m^{^{\prime }}\left( r\right) }{m\left( r\right) }\left( \frac{x_{j}}{r}%
\right) ,
\end{equation}%
and for $\xi _{1}=\xi _{2}=\xi /\sqrt{2}$, this would imply that%
\begin{equation}
2f\left( r\right) ^{2}=\left( \frac{m^{^{\prime }}\left( r\right) }{2m\left(
r\right) }\right) ^{2}\left( \frac{\xi _{1}^{2}x_{1}^{2}+\xi
_{2}^{2}x_{2}^{2}}{r^{2}}\right) \Longrightarrow f\left( r\right) =\frac{\xi
m^{^{\prime }}\left( r\right) }{4m\left( r\right) }.
\end{equation}%
Consequently, the corresponding two-dimensional PDM Euler-Lagrange equation
(17), for%
\begin{equation*}
m\left( r\right) =\frac{1}{1\pm \beta r^{2}},
\end{equation*}%
reads%
\begin{equation}
\frac{d}{dt}\left[ \frac{\dot{x}_{1}^{2}+\dot{x}_{2}^{2}+\omega ^{2}\xi ^{2}%
}{2(1\pm \beta r^{2})}\right] =0,
\end{equation}%
and admits solutions of the forms%
\begin{equation}
x_{j}\left( t\right) =A_{j}\cos \left( \Omega t+\varphi \right) \,;\,\Omega
^{2}=\frac{\mp \omega ^{2}\beta \xi ^{2}}{1\pm \beta \left(
A_{1}^{2}+A_{2}^{2}\right) }\,;\,\beta \geq 0;\,j=1,2\medskip \medskip
\end{equation}%
Under such settings, one can show that for $\beta =\mp 1/\xi ^{2}$ the
Lagrangian of (36) and the Lagrangian of (40) indulge the very same
dynamical properties as documented in the corresponding Euler-Lagrange
equation (16). Hence, the Lagrangian at hand here is a Mathews-Lakshmanan
type-II PDM-oscillators Lagrangian.

Obviously, moreover, our non-separable and non-superintegrable Lagrangian
(40) non-locally transforms into a separable and \emph{superintegrable}
Lagrangian (22). Our PDM Lagrangian (40) is a \emph{sub-superintegrable}
Mathews-Lakshmanan type-II PDM-oscillators Lagrangian, therefore. In short,
the relation%
\begin{equation}
\left\{ 
\begin{tabular}{c}
\ \emph{Superintegrable}\medskip \\ 
$L\left( q_{1},q_{2},\tilde{q}_{1},\tilde{q}_{2};\tau \right) \medskip $ \\ 
of \ (22) with\medskip \\ 
$q_{j}\left( \tau \right) =A_{j}\cos \left( \omega \tau +\varphi \right) $%
\end{tabular}%
\right\} \Longleftrightarrow \left\{ 
\begin{tabular}{l}
$\text{Nonlocal transformation\medskip }$ \\ 
$q_{j}=\xi _{j}\sqrt{m\left( r\right) }\medskip \,;\,j=1,2$ \\ 
$\tilde{q}_{j}=\dot{x}_{j}\sqrt{m\left( r\right) }\medskip $ \\ 
$f\left( r\right) =\frac{\xi m^{^{\prime }}\left( r\right) }{4m\left(
r\right) }\medskip ;\,\xi _{j}=\frac{\xi }{\sqrt{2}}$ \\ 
$m\left( r\right) =1/\left( 1\pm \beta r^{2}\right) ;\beta =\mp 1/\xi
^{2}\medskip $%
\end{tabular}%
\right\} \Longleftrightarrow \left\{ 
\begin{tabular}{c}
\emph{Sub-superintegrable}\medskip \\ 
$L\left( x_{1},x_{2},\dot{x}_{1},\dot{x}_{2};t\right) \medskip $ \\ 
of \ (39) with\medskip \\ 
$x_{j}\left( t\right) =A_{j}\cos \left( \Omega t+\varphi \right) \,\medskip
\medskip $ \\ 
$\text{ \ }\Omega ^{2}=\frac{\omega ^{2}}{1\pm \beta \left(
A_{1}^{2}+A_{2}^{2}\right) }\medskip $%
\end{tabular}%
\right\} .
\end{equation}%
represents the sought after mapping from/to \emph{superintegrable}
Lagrangian $L\left( q_{1},q_{2},\tilde{q}_{1},\tilde{q}_{2};\tau \right) $
of \ (22)\ to/from the \emph{Sub-superintegrable}\medskip\ Lagrangian $%
L\left( x_{1},x_{2},\dot{x}_{1},\dot{x}_{2};t\right) $ of \ (39).

\subsubsection{Sub-superintegrable Mathews-Lakshmanan type-III PDM
shifted-oscillators}

Let us now use a position-dependent mass with a different functional
structure moving in a shifted-oscillator force field and described by the
Lagrangian%
\begin{equation}
L\left( x_{1},x_{2},\dot{x}_{1},\dot{x}_{2};t\right) =\frac{1}{2}m\left(
r_{s}\right) \left\{ \dot{x}_{1}^{2}+\dot{x}_{2}^{2}-\omega ^{2}\left[
\left( x_{1}+\gamma _{1}\right) ^{2}+\left( x_{2}+\gamma _{2}\right) ^{2}%
\right] \right\} .
\end{equation}%
where $r_{s}=\sqrt{\left( x_{1}+\gamma _{1}\right) ^{2}+\left( x_{2}+\gamma
_{2}\right) ^{2}}$ is introduced for convenience. The form of the shifted
oscillator potential is clear here. At this point, we may recollect that the
functional nature/structure of the position-dependent mass $m\left( \bar{x}%
\right) $ determines the nature/structure of $g\left( \bar{x}\right) $ and $%
\,f\left( \bar{x}\right) $ in our nonlocal point transformation as suggested
by equation (19). Therefore, $f\left( r\right) \longrightarrow f\left(
r_{s}\right) $ and $g\left( r\right) \longrightarrow g\left( r_{s}\right) $
for our Lagrangian (46) at hand. Moreover, it is a straightforward manner,
and in parallel with (32)-(35), one may show that the substitution of%
\begin{equation}
q_{j}=\left( x_{j}+\xi _{j}\right) \sqrt{m\left( r_{s}\right) }%
\Longrightarrow f\left( r_{s}\right) =1+\frac{1}{4}\frac{m^{^{\prime
}}\left( r_{s}\right) }{m\left( r_{s}\right) }r_{s}
\end{equation}%
would consequently, for the choice 
\begin{equation}
f\left( r_{s}\right) =m\left( r_{s}\right) -\frac{1}{4}\frac{m^{^{\prime
}}\left( r_{s}\right) }{m\left( r_{s}\right) }r_{s}=1+\frac{1}{4}\frac{%
m^{^{\prime }}\left( r_{s}\right) }{m\left( r_{s}\right) }r_{s},
\end{equation}%
yield%
\begin{equation}
m\left( r_{s}\right) =\frac{1}{1\pm \beta r_{s}^{2}}=\frac{1}{1\pm \beta %
\left[ \left( x_{1}+\gamma _{1}\right) ^{2}+\left( x_{2}+\gamma _{2}\right)
^{2}\right] },
\end{equation}%
Then the corresponding two-dimensional PDM Lagrangian of (46) reads a
two-dimensional Mathews-Lakshmanan type-III PDM shifted-oscillators
Lagrangian%
\begin{equation}
L\left( x_{1},x_{2},\dot{x}_{1},\dot{x}_{2};t\right) =\frac{1}{2}\frac{\left[
\dot{x}_{1}^{2}+\dot{x}_{2}^{2}-\omega ^{2}r_{s}^{2}\right] }{1\pm \beta
r_{s}^{2}}.
\end{equation}%
Our two-dimensional PDM Lagrangian $L\left( x_{1},x_{2},\dot{x}_{1},\dot{x}%
_{2};t\right) $ (50) nonlocaly transforms into a \emph{superintegrable}
two-dimensional Lagrangian $L\left( q_{1},q_{2},\tilde{q}_{1},\tilde{q}%
_{2};\tau \right) $ of (22). Hence our PDM-Lagrangian $L\left( x_{1},x_{2},%
\dot{x}_{1},\dot{x}_{2};t\right) $ (50) is a \emph{sub-superintegrable }%
Lagrangian and the corresponding Euler-Lagrange equation (16) admits exact
solutions of the form%
\begin{equation}
x_{j}\left( t\right) =A_{j}\cos \left( \Omega t+\varphi \right) -\gamma
_{j};\,\Omega ^{2}=\frac{\omega ^{2}}{1\pm \beta \left(
A_{1}^{2}+A_{2}^{2}\right) };\,j=1,2.
\end{equation}%
The process is summed up as%
\begin{equation}
\left\{ 
\begin{tabular}{c}
\ Superintegrable\medskip \\ 
$L\left( q_{1},q_{2},\tilde{q}_{1},\tilde{q}_{2};\tau \right) \medskip $ \\ 
of \ (22) with\medskip \\ 
$q_{j}\left( \tau \right) =A_{j}\cos \left( \omega \tau +\varphi \right) $%
\end{tabular}%
\right\} \Longleftrightarrow \left\{ 
\begin{tabular}{l}
$\text{Nonlocal transformation\medskip }$ \\ 
$q_{j}=x_{j}\sqrt{m\left( r_{s}\right) }\medskip \,;\,j=1,2$ \\ 
$\tilde{q}_{j}=\dot{x}_{j}\sqrt{m\left( r_{s}\right) }\medskip $ \\ 
$f\left( r_{s}\right) =m\left( r_{s}\right) -\frac{1}{4}\frac{m^{^{\prime
}}\left( r_{s}\right) }{m\left( r_{s}\right) }r_{s}\medskip \medskip $ \\ 
$m\left( r_{s}\right) =1/\left( 1\pm \beta r_{s}^{2}\right) \medskip $%
\end{tabular}%
\right\} \Longleftrightarrow \left\{ 
\begin{tabular}{c}
Sub-superintegrable\medskip \\ 
$L\left( x_{1},x_{2},\dot{x}_{1},\dot{x}_{2};t\right) \medskip $ \\ 
of \ (50) with\medskip \\ 
$x_{j}\left( t\right) =A_{j}\cos \left( \Omega t+\varphi \right) -\gamma
_{j}\,\medskip $ \\ 
$\text{ \ }\Omega ^{2}=\frac{\omega ^{2}}{1\pm \beta \left(
A_{1}^{2}+A_{2}^{2}\right) }\medskip $%
\end{tabular}%
\right\} ,
\end{equation}%
to represent the mapping from the \emph{superintegrability} harmonic
oscillators (22) into \emph{sub-superintegrability} of the above
Mathews-Lakshmanan type-III PDM shifted-oscillators (50).

\subsection{Superintegrable shifted-linear oscillators into
sub-superintegrable nonlinear PDM-oscillators}

Consider a \emph{unit mass }particle moving in the two-dimensional
shifted-oscillators potential 
\begin{equation}
V(q_{1},q_{2})=\frac{1}{2}\left[ \alpha _{1}^{2}\left( q_{1}+\eta
_{1}\right) ^{2}+\alpha _{2}^{2}\left( q_{2}+\eta _{2}\right) ^{2}\right]
\Longrightarrow V_{j}\left( q_{j}\right) =\frac{1}{2}\alpha _{j}^{2}\left(
q_{j}+\eta _{j}\right) ^{2};\,j=1,2,
\end{equation}%
in the generalized coordinates, with the constant shifts $\eta _{1},$ $\eta
_{2}\in 
\mathbb{R}
$. Then, the corresponding two-dimensional Lagrangian%
\begin{equation}
L\left( q_{1},q_{2},\tilde{q}_{1},\tilde{q}_{2};\tau \right) =\frac{1}{2}%
\left( \tilde{q}_{1}^{2}+\tilde{q}_{2}^{2}\right) -\frac{1}{2}\left[ \alpha
_{1}^{2}\left( q_{1}+\eta _{1}\right) ^{2}+\alpha _{2}^{2}\left( q_{2}+\eta
_{2}\right) ^{2}\right] ,
\end{equation}%
yields the Euler-Lagrange equations%
\begin{equation}
\frac{d}{d\tau }\tilde{q}_{j}+\alpha _{j}^{2}\left( q_{j}+\eta _{j}\right)
=0\Longrightarrow q_{j}\left( \tau \right) =A_{j}\cos \left( \alpha _{j}\tau
+\varphi \right) -\eta _{j};\,j=1,2,
\end{equation}%
with the initial conditions that $\,q_{j}\left( 0\right) =A_{j}-\eta _{j},\,%
\tilde{q}_{j}\left( 0\right) =0$, say. The separability of this Lagrangian
is obvious. However, the verification of the superintegrability of such a
Lagrangian follows (step-by-step) from the complex factorization recipe
(c.f., e.g., \cite{42,43,44}) by introducing the complex functions%
\begin{equation}
Q_{j}=\tilde{q}_{j}+i\alpha _{j}\left( q_{j}+\eta _{j}\right)
\Longrightarrow \frac{d}{d\tau }Q_{j}=i\alpha _{j}Q_{j};\,\,j=1,2.
\end{equation}%
Which would, in turn, suggest that 
\begin{equation}
Q_{jk}=Q_{j}^{\alpha _{k}}\left( Q_{k}^{\ast }\right) ^{\alpha
_{j}}\Longrightarrow \frac{d}{d\tau }Q_{jk}=i\left( \alpha _{j}\alpha
_{k}-\alpha _{k}\alpha _{j}\right) Q_{jk}=0.;\,\,j,k=1,2.
\end{equation}%
That is, the complex function $Q_{jk}$ represent \emph{complex }constants of
motion with vanishing deformed/rescaled-time evolution $\tau $, Yet, it is
an easy task to show that the two fundamental integrals $I_{1}$ and $I_{2}$
are given through the relation 
\begin{equation}
Q_{jj}=\tilde{q}_{j}^{2}+\alpha _{j}^{2}\left( q_{j}+\eta _{j}\right)
^{2}\Longrightarrow 2E_{1}=I_{1}=Q_{11},\,2E_{2}=I_{2}=Q_{22},
\end{equation}%
Whereas, for $\alpha _{1}=\alpha _{2}=\alpha _{\circ }$, for example,%
\begin{equation}
Q_{12}=\left[ \tilde{q}_{1}\tilde{q}_{2}+\alpha _{\circ }^{2}\left(
q_{1}+\eta _{1}\right) \left( q_{2}+\eta _{2}\right) \right] +i\alpha
_{\circ }\left[ \left( q_{1}+\eta _{1}\right) \tilde{q}_{2}-\left(
q_{2}+\eta _{2}\right) \tilde{q}_{1}\right] ,
\end{equation}%
which identifies two real integrals of motion $I_{3}$ and $I_{4}$ such that 
\begin{equation}
I_{3}=\func{Re}Q_{12}=\tilde{q}_{1}\tilde{q}_{2}+\alpha _{\circ }^{2}\left(
q_{1}+\eta _{1}\right) \left( q_{2}+\eta _{2}\right) \text{ \ ; \ \ }I_{4}=%
\func{Im}Q_{12}=\alpha _{\circ }\left[ \left( q_{1}+\eta _{1}\right) \tilde{q%
}_{2}-\left( q_{2}+\eta _{2}\right) \tilde{q}_{1}\right] .
\end{equation}%
Therefore, our two-dimensional Lagrangian (54) admits \emph{%
superintegrability} in the generalized coordinates $\left(
q_{1},q_{2}\right) $ and in the deformed/rescaled time $\tau $.

Next, under the nonlocal transformation (20), along with the substitutions%
\begin{equation}
q_{j}=x_{j}\sqrt{m\left( r\right) }-\eta _{j}\Longrightarrow \frac{dq_{j}}{%
dx_{j}}=\sqrt{m\left( r\right) }\left[ 1+\frac{m^{\prime }\left( r\right) }{%
2m\left( r\right) }\left( \frac{x_{j}^{2}}{r}\right) \right] \Longrightarrow
f\left( r\right) =1+\frac{1}{4}\frac{m^{^{\prime }}\left( r\right) }{m\left(
r\right) }r.
\end{equation}%
This result looks very much the same as that of $f\left( r\right) $ used in
(33). Thus it would, again, with the assumption that%
\begin{equation*}
f\left( r\right) =m\left( r\right) -\frac{1}{4}\frac{m^{^{\prime }}\left(
r\right) }{m\left( r\right) }r=1+\frac{1}{4}\frac{m^{^{\prime }}\left(
r\right) }{m\left( r\right) }r\Longrightarrow m\left( r\right) =\frac{1}{%
1\pm \beta r^{2}},
\end{equation*}%
lead to the \emph{sub-superintegrable }Mathews-Lakshmanan type-I
PDM-oscillator (36). The corresponding Euler-Lagrange equation (16) admits
exact solutions as those in (37). Then, the \emph{sub-superintegrability} of
our Mathews-Lakshmanan type-I PDM-Lagrangian (36) turns out to be a
consequence of the \emph{superintegrability} of the linear oscillators (22)
and/or the \emph{superintegrability} of the shifted-oscillators (54).
Likewise, the \emph{sub-superintegrable} and exact solvable
Mathews-Lakshmanan type-I PDM-Lagrangian (36) may very well be nonlocally
transformed into two \emph{superintegrable} Lagrangians, a \emph{%
superintegrable} linear oscillator (22) and a \emph{superintegrable}
shifted-oscillator (54). That is, the relation 
\begin{equation}
\left\{ 
\begin{tabular}{c}
\emph{\ Superintegrable}\medskip \\ 
$L\left( q_{1},q_{2},\tilde{q}_{1},\tilde{q}_{2};\tau \right) \medskip $ \\ 
of (54), with\medskip \\ 
$q_{j}\left( \tau \right) =A_{j}\cos \left( \alpha _{j}\tau +\varphi \right)
-\eta _{j}$%
\end{tabular}%
\right\} \Longleftrightarrow \left\{ 
\begin{tabular}{l}
$\text{Nonlocal transformation\medskip }$ \\ 
$q_{j}=x_{j}\sqrt{m\left( r\right) }-\eta _{j}\,;\smallskip \medskip
\,j=1,2, $ \\ 
$\tilde{q}_{j}=\dot{x}_{j}\sqrt{m\left( r\right) }\medskip $ \\ 
$f\left( r\right) =m\left( r\right) -\frac{1}{4}\frac{m^{^{\prime }}\left(
r\right) }{m\left( r\right) }r\medskip $ \\ 
$m\left( r\right) =1/\left( 1\pm \beta r^{2}\right) $%
\end{tabular}%
\right\} \Longleftrightarrow \left\{ 
\begin{tabular}{c}
\emph{Sub-superintegrable}\medskip \\ 
$L\left( x_{1},x_{2},\dot{x}_{1},\dot{x}_{2};t\right) \medskip $ \\ 
of \ (36), with\medskip \\ 
$x_{j}\left( t\right) =A_{j}\cos \left( \Omega t+\varphi \right) \,\medskip $
\\ 
$\text{ \ }\Omega ^{2}=\frac{\omega ^{2}}{1\pm \beta \left(
A_{1}^{2}+A_{2}^{2}\right) }\medskip $%
\end{tabular}%
\right\} .
\end{equation}%
would describe the mapping from the \emph{superintegrable} Lagrangians (54)
into the \emph{sub-superintegrable} ones of (36).

\subsection{Superintegrable Isotonic Oscillator into sub-superintegrable
PDM-deformed Isotonic oscillator}

A \emph{"unit mass"} particle moving in a two-dimensional isotonic
oscillator potential field%
\begin{equation}
V(q_{1},q_{2})=\frac{1}{2}\left( \omega _{1}^{2}q_{1}^{2}+\omega
_{2}^{2}q_{2}^{2}+\frac{\beta _{1}}{q_{1}^{2}}+\frac{\beta _{2}}{q_{2}^{2}}%
\right) \Longrightarrow V_{j}\left( q_{j}\right) =\frac{1}{2}\left( \omega
_{j}^{2}q_{j}^{2}+\frac{\beta _{j}}{q_{j}^{2}}\right) ,
\end{equation}%
where $\,\omega _{j}=n_{j}\omega _{\circ };\,\,j=1,2$, is described by the
Lagrangian%
\begin{equation}
L\left( q_{1},q_{2},\tilde{q}_{1},\tilde{q}_{2};\tau \right) =\frac{1}{2}%
\left( \tilde{q}_{1}^{2}+\tilde{q}_{2}^{2}\right) -\frac{1}{2}\left( \omega
_{1}^{2}q_{1}^{2}+\omega _{2}^{2}q_{2}^{2}+\frac{\beta _{1}}{q_{1}^{2}}+%
\frac{\beta _{2}}{q_{2}^{2}}\right) ,
\end{equation}%
which is known to be the \emph{superintegrable} Smorodisky-Winternitz type
Lagrangian (c.f., e.g., \cite{29,38,40,42,47}). Its \emph{superintegrability}
can be verified through the two complex substitutions%
\begin{equation}
Q_{j}=\left( \tilde{q}_{i}^{2}-\omega _{j}^{2}q_{j}^{2}+\frac{\beta _{j}}{%
q_{j}^{2}}\right) +2i\omega _{j}q_{j}\tilde{q}_{j}\Longrightarrow \frac{d}{%
d\tau }Q_{j}=2i\omega _{j}Q_{j},\smallskip \medskip \,\,j=1,2,
\end{equation}%
that satisfy (25) with $Q_{jk}$ representing complex constants of motion and
leads to more than two integrals of motion that manifest \emph{%
superintegrability}. \ Moreover, the corresponding Euler-Lagrange equations
of which read two Ermakov-Pinney's like equations%
\begin{equation}
\frac{d}{d\tau }\tilde{q}_{j}=-\omega _{j}^{2}q_{j}+\frac{\beta _{j}}{%
q_{j}^{3}},
\end{equation}%
with the corresponding exact solutions%
\begin{equation}
q_{j}=\sqrt{\frac{A_{j}}{\omega _{j}}\sin \left( \omega _{j}\tau +\delta
_{j}\right) }\Longrightarrow \beta _{j}=-A_{j}^{2};\,j=1,2.
\end{equation}

Yet, under the nonlocal transformation setting in (38) our \emph{%
superintegrable} Lagrangian $L\left( q_{1},q_{2},\tilde{q}_{1},\tilde{q}%
_{2};\tau \right) $ in (64) nonlocaly transforms into a \emph{%
sub-superintegrable} Smorodisky-Winternitz like PDM-oscillators Lagrangian%
\begin{equation}
L\left( x_{1},x_{2},\dot{x}_{1},\dot{x}_{2};t\right) =\frac{1}{2}\left\{ 
\frac{\dot{x}_{1}^{2}+\dot{x}_{2}^{2}-\left( \omega _{1}^{2}x_{1}^{2}+\omega
_{2}^{2}x_{2}^{2}\right) }{1\pm \lambda \left( x_{1}^{2}+x_{2}^{2}\right) }-%
\left[ 1\pm \lambda \left( x_{1}^{2}+x_{2}^{2}\right) \right] \left( \frac{%
\beta _{1}}{x_{1}^{2}}+\frac{\beta _{2}}{x_{2}^{2}}\right) \right\} .
\end{equation}%
The Euler-Lagrange equation (16) for which admits exact solutions of the form%
\begin{equation}
x_{j}=\sqrt{\frac{A_{j}}{\Omega }\sin \left( \Omega \tau +\delta _{j}\right) 
};\,j=1,2\medskip \medskip ,
\end{equation}%
where%
\begin{equation}
\Omega ^{2}=\left\{ 
\begin{tabular}{ll}
$\omega ^{2}+\lambda ^{2}\left( A_{1}+A_{2}\right) ^{2}$ & $;$ \medskip\ $%
\omega _{1}=\omega _{2}=\omega ,\beta _{1}\neq \beta _{2}$ \\ 
$\frac{A_{1}\omega _{1}^{2}+A_{2}\omega _{2}^{2}+\lambda ^{2}\left(
A_{1}+A_{2}\right) ^{3}}{A_{1}+A_{2}}\medskip $ & $;$ $\ \omega _{1}\neq
\omega _{2},\beta _{1}\neq \beta _{2}$ \\ 
$\left( \omega _{1}^{2}+\omega _{2}^{2}\right) /\left( 16A^{2}\lambda
^{2}\right) \medskip $ & $;$ \ $\omega _{1}\neq \omega _{2},\beta _{1}=\beta
_{2}\Leftrightarrow A_{1}=A_{2}=A$%
\end{tabular}%
\right. .
\end{equation}

\section{Concluding Remarks}

In this article, and in parallel with our recent methodical proposal in \cite%
{38}, we have introduced the two-dimensional extension of the
one-dimensional PDM-Lagrangians and their nonlocal transformation mappings'
recipes into constant \emph{unit-mass} exactly solvable Lagrangians. Hereby,
the two-dimensional nonlocal point transformations (10) and the related
Euler-Lagrange equations invariance conditions (19) are reported. However,
dealing with Lagrangians in more than one-dimension renders \emph{%
superintegrability} to be unavoidably in the process. We have, therefore,
asserted that, if a set of \emph{"superintegrable
reference/target-Lagrangians"} $L\left( q_{1},q_{2},\tilde{q}_{1},\tilde{q}%
_{2};\tau \right) $ is mapped into a set of \emph{"PDM
target/reference-Lagrangians"} $L\left( x_{1},x_{2},\dot{x}_{1},\dot{x}%
_{2};t\right) $, then the set of \emph{"PDM target/reference-Lagrangians"} $%
L\left( x_{1},x_{2},\dot{x}_{1},\dot{x}_{2};t\right) $\ is a set of \emph{%
"sub-superintegrable PDM-Lagrangians"} (as shall be so labelled
hereinafter). Two sets of illustrative examples are used. The first set is
devoted to the mappings from \emph{superintegrable} linear oscillators into 
\emph{sub-superintegrable} nonlinear PDM-oscillators. Where, three
two-dimensional \emph{sub-superintegrable} Mathews-Lakshmanan type
PDM-oscillators are used: (i) a \emph{sub-superintegrable}
Mathews-Lakshmanan type-I PDM-oscillator for a PDM-particle moving in a
harmonic oscillator potential, (ii) a \emph{sub-superintegrable}
Mathews-Lakshmanan type-II PDM-oscillator for a PDM-particle moving in a
constant potential, and (iii) a \emph{sub-superintegrable} shifted
Mathews-Lakshmanan type-III PDM-oscillator for a PDM-particle moving in a
shifted harmonic oscillator potential. The second set, nevertheless,
includes some \emph{superintegrable} shifted linear oscillators (new to the
best of our knowledge) and isotonic oscillators that are mapped into \emph{%
sub-superintegrable} PDM-nonlinear and \emph{sub-superintegrable}
PDM-isotonic oscillators, respectively. The mappings included exact
solvability as well. Our observations are in order.

Whilst the two-dimensional PDM Mathews-Lakshmanan type-I and type-III, and
the PDM shifted nonlinear oscillators share the same total energy%
\begin{equation}
E_{tot}=\frac{1}{2}\omega ^{2}\frac{\left( A_{1}^{2}+A_{2}^{2}\right) }{1\pm
\beta \left( A_{1}^{2}+A_{2}^{2}\right) },
\end{equation}%
the two-dimensional PDM Mathews-Lakshmanan type-II oscillators admit total
energy%
\begin{equation}
E_{tot}=\frac{\omega ^{2}\xi ^{2}}{1-\xi ^{2}\left(
A_{1}^{2}+A_{2}^{2}\right) },
\end{equation}%
and the two-dimensional PDM-deformed isotonic oscillators indulge total
energy%
\begin{equation}
E_{tot}=\frac{1}{2}\left\{ \frac{\omega _{1}^{2}A_{1}+\omega _{2}^{2}A_{2}}{%
\Omega \pm \lambda \left( A_{1}+A_{2}\right) }-\left[ \Omega \pm \lambda
\left( A_{1}+A_{2}\right) \right] \left( A_{1}+A_{2}\right) \right\} .
\end{equation}%
Yet, the two-dimensional PDM Mathews-Lakshmanan type-I and type-III inherit
the dynamical properties and trajectories of each other. On the other hand,
the \emph{sub-superintegrability} of the PDM Mathews-Lakshmanan type-I
oscillators (36) may very well be attributed to the \emph{superintegrability}
of the linear oscillator (22) and/or the shifted-oscillators (54).

Finally, the generalization of the current methodical proposal into more
than two-dimensional recipes looks eminent and feasible. This is very
obviously documented in the description of our equations (6) to (12), where $%
j=1,2$ is used. Strictly speaking, for a three-dimensional case $%
j=1,2\rightarrow $ $j=1,2,3$ and $\bar{x}=x_{1},x_{2}\rightarrow \bar{x}%
=x_{1},x_{2},x_{3}$ in (10) and so on so forth. It would be interesting to
study and explore the consequences of such generalization.


\newpage

\begin{thebibliography}{99}
\bibitem{1} O. von Roos, Phys. Rev. \textbf{B 27 }(1983) 7547.

\bibitem{2} A. de Souza Dutra, C A S Almeida, Phys Lett. \textbf{A 275}
(2000) 25.

\bibitem{3} S. H. Mazharimousavi, J Phys \textbf{A}: Math. Theor.\textbf{41 (%
}2008\textbf{) }244016.

\bibitem{4} O. Mustafa, S. H. Mazharimousavi, Phys. Lett. \textbf{A 358}
(2006) 259.

\bibitem{5} A. D. Alhaidari, Phys. Rev. \textbf{A 66} (2002) 042116.

\bibitem{6} B. Bagchi, A. Banerjee, C. Quesne, V. M. Tkachuk, J. \ Phys. 
\textbf{A}: Math. Gen. \textbf{38} (2005) 2929.

\bibitem{7} O. Mustafa, S. H. Mazharimousavi, Phys. Lett. \textbf{A 357}
(2006) 295.

\bibitem{8} O. Mustafa, J Phys \textbf{A}: Math. Theor. \textbf{44 (}2011%
\textbf{) }355303.

\bibitem{9} B. Bagchi, P. Gorain, C. Quesne and R. Roychoudhury, Mod. Phys.
Lett. \textbf{A 19} (2004) 2765.

\bibitem{10} O. Mustafa, S. H. Mazharimousavi, Int. J. Theor. Phys \ \textbf{%
46} (2007) 1786.

\bibitem{11} S. Cruz y Cruz, O Rosas-Ortiz, J Phys \textbf{A}: Math. Theor. 
\textbf{42} (2009) 185205.

\bibitem{12} S. Cruz y Cruz, J. Negro and L.M. Nieto, Phys. Lett. \textbf{A} 
\textbf{369,} (2007) 400.

\bibitem{13} S. Cruz y Cruz, J. Negro and L.M. Nieto, Journal of Physics:
Conference Series \textbf{128, }(2008)\textbf{\ }012053.

\bibitem{14} S. Cruz y Cruz, O Rosas-Ortiz, SIGMA \textbf{9, }(2013)\textbf{%
\ }004.

\bibitem{15} S. Ghosh and S. K. Modak, Phys. Lett. A \textbf{373,} (2009)
1212.

\bibitem{16} B. Bagchi, S. Das, S. Ghosh and S. Poria, J. Phys. \textbf{A}:
Math. Theor. \textbf{46,} (2013) 032001.

\bibitem{17} S. H. Mazharimousevi, O. Mustafa, Phys. Scr. \textbf{87} (2013)
055008.

O. Mustafa; arXiv:1208.2109:\textbf{\ }Comment on the "Classical and quantum
position-dependent mass harmonic oscillators" and ordering-ambiguity
resolution.

\bibitem{18} P. M. Mathews, M. Lakshmanan, Quart. Appl. Math. \textbf{32 }%
(1974)\textbf{\ }215.

\bibitem{19} A. Venkatesan, M. Lakshmanan, Phys. Rev. \textbf{E} \textbf{55}
(1997) 5134.

\bibitem{20} J. F. Cari\~{n}ena, M. F. Ra\~{n}ada, M. Santander, Regul.
Chaotic Dyn. \textbf{10} (2005) 423.

\bibitem{21} A. Bhuvaneswari, V. K. Chandrasekar, M. Santhilvelan, M.
Lakshmanan, J. Math. Phys. \textbf{53} (2012) 073504.

\bibitem{22} O. Mustafa, J. Phys. A: Math. Theor.\textbf{\ 46 }(2013) 368001.

\bibitem{23} A. K. Tiwari, S. N. Pandey, M. Santhilvelan, M. Lakshmanan, J.
Math. Phys. \textbf{54} (2013) 053506.

\bibitem{24} M. Lakshmanan, V. K. Chandrasekar, Eur. Phys J. ST \textbf{222}
(2013) 665.

\bibitem{25} Z. E. Musielak, J. Phys. \textbf{A}: Math. Theor. \textbf{41,}
(2008) 055205.

\bibitem{26} C. Quesne, J. Math. Phys. \textbf{56} (2015) 012903.

\bibitem{27} C. Quesne, V. M. Tkachuk, J. Phys. \textbf{A 37} (2004) 4267.

\bibitem{28} B. Bagchi, A. Banerjee, C. Quesne, V. M. Tkachuk, J. Phys. 
\textbf{A 38} (2005) 2929.

\bibitem{29} J. F. Cari\~{n}ena, M. F. Ra\~{n}ada, M. Santander, SIGMA 
\textbf{3} (2007) 030.

\bibitem{30} J. F. Cari\~{n}ena, M. F. Ra\~{n}ada, M. Santander, Ann. Phys. 
\textbf{322 }(2007) 434.

\bibitem{31} J. F. Cari\~{n}ena, M. F. Ra\~{n}ada, M. Santander, Rep. Math.
Phys. \textbf{54} (2004) 285.

\bibitem{32} C. Muriel, J. L. Romero, J. Phys. \textbf{A}: Math. Theor. 
\textbf{43} (2010) 434025.

\bibitem{33} R. G. Pradeep, V. K. Chandrasekar, M. Santhilvelan, M.
Lakshmanan, J. Math. Phys. \textbf{50} (2009) 052901.

\bibitem{34} N. Euler, M. Euler, J. Nonlinear Math. Phys. \textbf{11} (2004)
399.

\bibitem{35} W. -H. Steeb, "\textit{Invertible Point Transformation and
Nonlinear Differential Equations", }World Scientific, Singapore, 1993.

\bibitem{36} K. S. Govinder, P. G. L. Leach, J. Math. Anal. Appl. \textbf{%
287 }(2003) 399.

\bibitem{37} N. Euler, T. Wolf, P. G. L. Leach, M. Euler, Acta Appl. Math. 
\textbf{76} (2003) 89.

\bibitem{38} O. Mustafa, J. Phys. \textbf{A}; Math. Theor. \textbf{48}
(2015) 225206.

\bibitem{39} M. Ranada, M. A. Rodrigues, and M Santander, J. Math. Phys. 
\textbf{51}, 042901 (2010).

\bibitem{40} M. A. Rodrigues, p. Tempesta, and P. Winternitz, Phys. Rev. E 
\textbf{78} 046608 (2008).

\bibitem{41} N. W. Evans and P. N. Verrier, J. Math. Phys. \textbf{49},
092902 (2008).

\bibitem{42} M. Ranada, J. Math. Phys. \textbf{57}, 052703 (2016).

\bibitem{43} J. M. Jauch and E. L. Hill. Phys. Rev. \textbf{57}, 641 (1940).

\bibitem{44} A. M. Perelomov, \emph{integrable systems of classical
mechanics and Lie algebras} (Birkhauser, Base, 1990).

\bibitem{45} V. Ovsienko, R. E. Schwartz, S.Tabachnikov arXiv:1107.3633:
Liouville-Arnold integrability of the pentagram map on closed polygons.

\bibitem{46} M. Ranada, J. Math. Phys. \textbf{38}, 4165 (1997).

\bibitem{47} F. Tremblay, A. Turbiner, P. Winternitz, J. Phys. \textbf{A};
Math. Theor. \textbf{43} (2010) 015202; J. Phys. A \textbf{43}: Math. Theor.
(2010) 015202,

\bibitem{48} W. Miller, S. Post, P. Winternitz, J. Phys. \textbf{A}; Math.
Theor. \textbf{46} (2013) 423001

\bibitem{49} R. C.-Stursberg, J. Math. Phys. \textbf{55} (2014) 042904.

\bibitem{50} M. Ranada, Phys. Lett. \textbf{A 379} (2015) 2267.

\bibitem{51} J. F. Cari\~{n}ena, F. J. Herranz, M. F. Ra\~{n}ada, J. Math.
Phys. \textbf{58} (2017) 022701.
\end{thebibliography}
\end{document}